# Systematic study of defect-related quenching of NV luminescence in diamond with time correlated single photon counting spectroscopy

D. Gatto Monticone[1,2,3,6], F. Quercioli[4], R. Mercatelli[4], S. Soria[5],

S. Borini[7,#], T. Poli[8], M. Vannoni[4,§], E. Vittone[1,2,3,6], P. Olivero[1,2,3,6]*

*[1] Physics Department - University of Torino, Torino, Italy*

*[2] NIS Centre of Excellence - University of Torino, Torino, Italy*

*[3] INFN Sez. Torino, Torino, Italy*

*[4] CNR - Istituto Nazionale di Ottica, Firenze, Italy*

*[5] CNR - Istituto di Fisica Applicata "Nello Carrara"", Firenze, Italy*

*[6] Consorzio Nazionale Interuniversitario per le Scienze Fisiche della Materia (CNISM) - sez. Torino, Italy*

*[7] Thermodynamics Division - Istituto Nazionale di Ricerca Metrologica (INRiM), Torino, Italy*

*[8] Chemistry Department - University of Torino, Italy*

**Abstract**

We report on the systematic characterization of photoluminescence (PL) lifetimes in $NV^-$ and $NV^0$ centers in 2 MeV $H^+$ implanted type Ib diamond samples by means of a timecorrelated single photon counting (TCSPC) microscopy technique. A dipole-dipole resonant energy transfer model was applied to interpret the experimental results, allowing a quantitative correlation of the concentration of both native (single substitutional nitrogen atoms) and ion-induced (isolated vacancies) PL-quenching defects with the measured PL lifetimes. The TCSPC measurements were carried out in both frontal (i.e. laser beam probing the main sample surface along the same normal direction of the previously implanted ions) and lateral (i.e. laser beam probing the lateral sample surface orthogonally with respect to the same ion implantation direction) geometries. In particular, the latter geometry allowed a direct probing of the centers lifetime along the strongly nonuniform damage profiles of MeV ions in the crystal. The extrapolation of empirical quasi-exponential decay parameters allowed the systematic estimation of the mean quantum efficiency of the centers as a function of intrinsic and ion-induced defect concentration, which is of direct relevance for the current studies on the use of diamond color centers for photonic applications.

\* corresponding author:    paolo.olivero@unito.it

ph: +39 011 670 7366

fax: +39 011 670 7020

present affiliations:    # Nokia Research Center, Cambridge (UK)

§ European XFEL GmbH, Hamburg, Germany

# 1. Introduction

The study of negatively-charged single nitrogen-vacancy (NV⁻) luminescent centers in diamond attracted a growing interest in the last decades, due to the opportunities they offer in the coherent manipulation of quantum states at room temperature, as well as in the efficient and high-rate emission of singlephotons on demand. Such unique properties make this center appealing not only in fundamental quantum optics [1-3], but also in advanced applications such as quantum computing [4-6], singlespin-based magnetic, electrical and biological sensing [7-9], quantum cryptography [10-12] and quantum nanomechanics [13-15].

A key advantage of single luminescent centers in diamond such as the NV⁻ complex is based on the fact that, since they are usually consisting of deep defects in a wide bandgap material, they can be suitably considered as the solid-state analogous of trapped atoms inside a spin-free environment characterized by a broad optical transparency. This is of course only true in an ideal crystal, and several works were devoted in assessing the variation of the spectral and spincoherence properties of the NV⁻ center depending on the concentration in the hosting crystal of structural defects [16, 17], isotopic $^{13}$C impurities [18-20] and foreign substitutional atoms such as nitrogen [21, 22].

The important issue of achieving single NV⁻ centers in bulk single-crystal diamond with minimum interaction with the surrounding crystal was addressed in a series of works with two approaches: optimizing the ion-implantation and post-implantation processing [23-30] and manipulating the NV⁻ centers with suitable noise-correcting procedures [31-37].

Luminescence lifetime is an effective tool to directly study the non-radiative decay channels in colour centers, thus giving significant information on their quantum efficiency, as much as on their interaction with the surrounding crystal environment [38, 39]. Luminescence lifetime studies have been applied in the characterization of defect interactions for various luminescent centers in diamond, such as the H3 [40, 41] and N3 [42] centers. In both of the above-cited works a significant decrease in lifetime was observed in samples characterized by high nitrogen concentrations.

In early works, the lifetime of $NV^-$ centers was evaluated as $(13 \pm 0.5)$ ns, although a relation between the centers lifetime and the crystal quality in synthetic samples was identified [43, 44].

The $NV^-$ center is characterized by a triplet ($S = 1$) spin state, with different spin-projection states ($m_S = 0$ and $m_S = \pm 1$) being characterized by the same oscillator strength. However, intersystem crossing processes involving non-radiative transitions through intermediate metastable singlet states determine different lifetimes (13.7 ns and 7.3 ns, respectively [45]) for the $m_S = 0$ and $m_S = \pm 1$ spin-projection states of the defect, as found in lifetime measurements combined with MW manipulation of the excited states [45, 46].

Photoluminescence lifetime mapping measurements with a time-correlated single photon counting (TCSPC) technique proved to be an effective tool to investigate the quenching mechanisms of $NV^-$ and $NV^0$ centers [40, 47].

In the present work we report a systematic study on the variation of the lifetime of NV (i.e. $NV^-$ and $NV^0$) centers as a function of ion-induced damage density in artificial HPHT diamonds with different nitrogen concentrations, performed with a time-correlated single-photon-counting micro-spectroscopy technique.

## 2. Experimental

*2a. Samples preparation and preliminary FTIR characterization*

In the present study three single-crystal type Ib diamond samples produced by Element Six with the High Pressure High Temperature (HPHT) technique were employed. The samples size was 3×3×0.3 mm$^3$. While samples #1 and #2 were optically polished on one of their two larger faces, sample #3 was optically polished on both frontal and lateral faces. Samples with a single growth sector were chosen, as observed by cross-polarization optical microscopy and infrared microscopy. The above-mentioned growth macro-sectors develop during the HPHT synthesis from a single seed and, depending on their orientation, they are characterized by different impurity concentrations [48-50]. Thus, it was assumed that the absence of observable sectors resulted in a uniform distribution of impurities within the sample.

The substitutional nitrogen concentration in the samples was characterized with micro-Fourier Transform InfraRed absorption (µ-FTIR) transmission measurements, using a Bruker Vertex 70 FTIR IR spectrometer coupled to an IR optical microscope Bruker Hyperion 3000 equipped with an Infrared Associates Inc. MCT detector in the spectral range from 4000 to 600 cm$^{-1}$ with an average spectral resolution of 4 cm$^{-1}$.

In Fig. 1 typical µ-FTIR spectra obtained from samples #1, #2 and #3 are reported. Sample #1 is characterized by the lowest nitrogen concentration, therefore the absorption feature at wavenumber $1/\lambda = 1130$ cm$^1$ (indicated by the black arrow) is less pronounced. On the other hand, samples #2 and #3 are characterized by similar nitrogen concentrations. The assumption on samples homogeneity was qualitatively confirmed by probing the samples at 10 different 20×20 µm$^2$ locations (each close to one of the implanted areas), and ~10% variations on measured spectra were observed. The absorption coefficient at $1/\lambda = 1130$ cm$^{-1}$ was measured to estimate the concentration of single substitutional nitrogen ($[N_S]$), adopting the calibration reported in [51]. The resulting estimates of $[N_S]$ are $(0.60 \pm 0.05) \times 10^2$ ppm, $(2.0 \pm 0.2) \times 10^2$ ppm and $(1.9 \pm 0.3) \times 10^2$ ppm for samples #1, #2 and #3, respectively.

*2b. Ion implantation and post-processing*

Samples were implanted at room temperature across the polished surface with 2 MeV H$^+$ ions at the AN2000 microbeam facility of the Legnaro National Laboratories with a rasterscanning ion microbeam focused to a size of ~5 µm, in order to deliver a uniform fluence across the irradiated areas. While samples #1 and #2 were implanted across the frontal surface in 100×100 µm$^2$ and 200×200 µm$^2$ square areas at different fluences ranging from $5 \times 10^{14}$ cm$^{-2}$ to $1 \times 10^{17}$ cm$^{-2}$ (from now on, these are referred as "frontal implantations"), sample #3 was implanted in two areas across its edge at fluences $1 \times 10^{15}$ cm$^{-2}$ and $2 \times 10^{17}$ cm$^{-2}$, with the purpose of performing crosssectional optical characterization in a lateral geometry, as schematically shown in Fig. 2 (from now on, these are referred as "lateral implantations"). In all implantations beam current was ~0.5 nA. Fig. 3 shows the strongly non-uniform depth profile of the linear damage density, as evaluated with the SRIM2008.04 Monte Carlo code [52] in "Detailed calculation with full damage cascade" mode by taking an atom displacement energy value of 50 eV [53].

After ion implantation, the samples were thermally annealed in vacuum ($p<10^{-4}$ Pa) for two hours at a temperature of 800 °C, which is considered suitable for the conversion to NV centers of a large fraction of the ion-induced vacancies and of the native nitrogen [54].

*2c. PL characterization*

Room temperature photoluminescence (PL) spectroscopy was performed with the purpose of preliminarily assessing the spectral features of the induced NV luminescence in the samples. A Jobin Yvon Raman micro-spectrometer was employed for this scope, with 532 nm laser excitation and a CCD Andor "DU420A-OE" detector. Subsequent PL spectra from each sample are mutually comparable in terms of absolute PL intensities within a reasonable level of confidence, since particular care was taken in carrying the measurements within a reasonably short timeframe in the same experimental conditions, particularly as regards the surface focusing procedure. Figs. 4a and 4b show the PL spectra collected from the different implanted areas of samples #1 and #2, respectively. In the spectra, the $NV^-$ (zero phonon line – ZPL: $\lambda = 638$ nm) and $NV^0$ (ZPL: $\lambda = 575$ nm) emissions are visible together with their respective phonon sidebands, while no emission related to single vacancies (ZPLs: $\lambda = 740.9$ nm, 744.4 nm; also referred as "GR1") is visible. The latter evidence provides a quantitative indication that after 800 °C annealing a large fraction of induced vacancies recombined with native nitrogen to form NV complexes.

The PL emission from sample #1 is characterized by a lower NV$^-$:NV$^0$ emission ratio with respect to sample #2, in which NV$^0$ emission is very weak. This can be explained if it is considered that in low-nitrogen samples the NV centers are less likely to be in proximity with nearby donors, and therefore are more probably found in their neutral charge state; on the other hand, NV$^-$ emission is predominant in samples characterized by high nitrogen concentration [55]. Also, it is worth noting that, compatibly with previous reports [16], the evolution of the overall PL intensity as a function of implantation fluence follows a non-monotonic trend. This can be explained by considering the effect of the increasing defect concentration induced by higher fluence implantations. At low fluences this causes an initial increase of PL yield as more and more luminescent centers are generated, while at high fluences an increasing defect density starts being detrimental to the PL yield, due to the combined effects of damage-induced optical absorption and non-radiative coupling with the existing luminescent centers (i.e. quenching). The latter process is systematically investigated in the present work by means of TCSPC technique.

The intensities of the NV$^-$ and NV$^0$ ZPL emissions were evaluated by spectral integration after suitable background subtraction, and in Fig. 5 the evolution as a function of implantation fluence of the (NV$^-$:NV$^0$) ratio is reported. At increasing fluences, the (NV$^-$:NV$^0$) ratio exhibits a monotonically decreasing trend in both samples. This observation is compatible with previous reports [16] and has several possible explanations, which are not mutually exclusive: i) a decrease in the concentration of negatively-charged NV centers due to the lowering of the Fermi level caused by the conversion of native nitrogen atoms into NV complexes, ii) a decrease in the concentration of negatively-charged NV centers due to the electron-trapping effect of induced defects and iii) a more effective nonradiative coupling of induced defects with the $NV^-$ centers with respect to the $NV^0$ centers.

*2d. Time-correlated single-photon counting (TCSPC) measurements*

In the TCSPC measurements the luminescence excitation was provided by a tunable Ti:Sapphire modelocked laser source (700-890 nm, Coherent Mira 900 F) pumped by a frequency-doubled Nd:YVO$_4$ laser emitting at 532 nm (5W CW, Coherent Verdi V5). The laser emission was modulated with a cavity dumper (APE Berlin Pulse Switch) in order to obtain a pulse repetition rate of 10 MHz. The pulse duration is of the order of $10^2$ fs. The excitation laser light was sent into a microstructured fibre (NL-1.7-670 crystal fibre) for supercontinuum generation and the excitation band was selected by a $(485 \pm 12)$ nm bandpass filter (HQ485/25 M, Chroma Technology Corporation). The excitation light was then coupled into a Nikon PCM2000 Confocal Laser Scanning Microscope (CLSM) unit equipped with a Nikon TE2000-U inverted optical microscope. In all measurements, the microscope was equipped with a Nikon 40× Plan Fluor objective (numerical aperture NA = 0.75, air), with the exception of the measurements performed on sample #3, in which a Nikon 60× Plan Fluor (NA = 1.4, oil) was employed with no coverslip between the lens and the sample. The microscopy system had a typical lateral spatial resolution of ~1 μm while the probe depth of the system was estimated to be ~10 μm. In Fig. 3 the probing depth is highlighted, showing that in frontally-implanted samples a constant damage density is probed.

The fluorescence light was collected into a single-grating spectrometer (Oriel Fics 77442) equipped with a 405 grooves/mm grating, and then detected with a Becker & Hickl PML-16 multichannel head containing a 16-channels photomultiplier tube (Hamamatsu R5900-L16) and the relevant electronics. The signal is then processed with the TCSPC technique that allows the measurement of the fluorescence decay curve for each pixel of the acquired image. The core element of the TCSPC system is the SPC-830 module installed on a single PC board. The TCSPC technique is based on the measurement of the arrival times of individual photons with respect to the excitation pulse and on the subsequent statistical reconstruction of the decay curve from multiple single-photon measurements, on a pixel-by-pixel basis [56]. The time window of the acquired chronograms is 100 ns, with a 256 time-bin resolution.

From a spectral point of view, it is possible to collect the fluorescence in the $\lambda = 498 \div 738$ nm range, which is subdivided into 16 spectral channels. In the present work, the $(730 \pm 8)$ nm and the $(595 \pm 8)$ nm spectral intervals were used to acquire respectively the PL signal from the phonon sidebands of the $NV^-$ and $NV^0$ emissions, as shown in Fig. 4a. The collection of the PL signals at the phonon sidebands allowed a significant increase in signal statistics, given the fact that, due to strong phonon coupling, in NV centers only a small fraction of the PL emission is concentrated in the ZPL. More importantly, the $(730 \pm 8)$ nm spectral interval for $NV^-$ detection was chosen to avoid spectral overlap with the tail of the $NV^0$ phonon sidebands, while retaining significant signal intensity from the phonon sideband of the $NV^-$ emission. In a series of preliminary tests, wherever possible (i.e. where no significant spectral overlap occurred) it was verified that the lifetime behavior of ZPL and phonon-sideband emissions are mutually consistent, for both kind of centers.

In general, PL decay chronograms relevant to regions implanted at different fluences in samples #1 and #2 were acquired by grouping the signals acquired from different pixels within each implanted region, in order to improve signal statistics. The Gaussian shape of the scanning ion microbeam employed to carry the implantation determines a lack of sharpness at the edges of the irradiated regions, therefore care was taken in avoiding pixels too close to the edges of the implanted regions (see Fig. 6a).

The temporal response function of the instrument (defined as a convolution of the finite duration of the excitation pulse and of the time resolution of the acquisition system) was determined by measuring the duration of the excitation pulse which was leaking through the monochromator at the $(506 \pm 8)$ nm spectral channel, resulting in a Gaussian pulse with a FWHM of 780 ps, corresponding to 2 of the 256 time-bins into which the 100 ns temporal window was subdivided.

## 3. Theory

It is well known that the luminescent emission of an unperturbed excited photo-active system (such as an ideal single NV center in a perfect diamond crystal) has an exponential time dependence characterized by a specific decay rate. The presence of a nearby system (such as a structural defect or an impurity in the diamond crystal, generally referred as a "quencher" of the optical transition under investigation) to which the luminescent center is resonantly coupled determines the appearance of a new non-radiative decay channel for the center. Therefore, the photon emission rate $I(t)$ is defined in this case as:

$$I(t)=\frac{-dN(t)}{dt}=a \cdot N_0 \cdot \exp[-(a+K) \cdot t] \tag{1}$$

where $N$ is the number of excited luminescent centers, $a$ is the intrinsic radiative decay rate of the luminescent center and $K$ is the non-radiative decay rate associated with the resonant coupling with the nearby defect/impurity. The latter term depends from the strength of the resonant coupling, which in turn depends from the distance between the two systems. A dipole-dipole coupling determines a dependence from the 6$^{th}$ power of the distance, while a dipole-quadrupole coupling determines a dependence from the 8$^{th}$ power of the distance, and so on.

In a more realistic case, the luminescent center is coupled to a large number of nearby defects/impurities located at different distances. According to the theory developed in [57, 38, 39], an integration of the different resonant coupling strengths for randomly located quenchers around the luminescent center results in the following deformed exponential decay rate:

$$I(t)=a \cdot N_0 \cdot \exp(-a \cdot t) \cdot \exp[-k \cdot (a \cdot t)^c] \tag{2}$$

where $k$ is now an a-dimensional parameter dependent from the quenchers concentration and the intensity of the resonant coupling, while the value of $c$ depends on the nature of the resonant coupling ($c = ½$ for dipole-dipole interaction, $c = ⅜$ for dipole-quadrupole interaction).

The mean quantum efficiency $\eta$ of luminescent centers surrounded by a given quencher distribution can be derived as the ratio between its total number of radiative decays (obtained by integrating Eq. 2) and the total number of radiative decays of a corresponding isolated center (obtained by integrating a simple exponential function). Therefore, in the case of dipole-dipole coupling, the quantum efficiency $\eta$ can be derived as a function of the "coupling strength" parameter $k$ as follows [38]:

$$\eta(k) = \frac{1}{2} \cdot \left\{ 2 + \sqrt{\pi} \cdot k \cdot \exp\left(\frac{k^2}{4}\right) \cdot \left[ erf\left(\frac{k}{2}\right) - 1 \right] \right\} \quad (3)$$

where $erf$ is the Gaussian error function.

With more specific reference to our system of interest (i.e. NV centers in a defect/impurities containing diamond lattice), some additional preliminary considerations can be formulated on the basis of the information available in literature.

Substitutional nitrogen defects behave as quenchers of the NV emission, as much as of other luminescent centers [41, 42]. As mentioned above, evidence of dipoledipole resonant energy transfer from $NV^-$ centers to single substitutional nitrogen in an artificial diamond was reported in early works [43, 44] and more recently investigated with the TCSPC technique [40]. In [40], also A aggregates (i.e. $N_2$ complexes) were identified as effective quenchers of $NV^-$ emission, while B aggregates (i.e. $N_4$ complexes) do not appear as effective in this respect. In a subsequent work, also the dipole-dipole resonant energy transfer from $NV^0$ centers to neutral substitutional nitrogen was investigated in detail [47].

Several previous works report extensively about the effect on the intensity of NV centers of structural defects induced from neutron [58], electron [59] and ion [16] irradiation, and it is natural to assume that also the centers lifetime is affected by radiation-induced damage. While a 2 MeV electron irradiation at a fluence of $2 \times 10^{18}$ cm$^{-2}$ does not seem to have significant effects on the centers lifetime [40], in the present work we report a systematic investigation on the effect of ion implantation on NV centers lifetime in samples characterized by different substitutional nitrogen concentrations.

## 4. Experimental results

*4a. Data analysis*

Figure 6a shows a typical TCSPC map acquired in the $(730 \pm 8)$ nm spectral window (corresponding to the phonon sideband of the NV$^-$ emission) from a 100×100 µm$^2$ region of sample #2 implanted with 2 MeV H$^+$ ions at a fluence of $1\times10^{16}$ cm$^{-2}$. The intensity map clearly highlights the implanted region where a higher concentration of NV centers is formed. It is worth stressing that, apart from PL intensity, each pixel encodes a PL decay chronogram. The typical region from which chronograms are extracted and analyzed is highlighted by the white dashed square.

Figure 6b shows chronograms of normalized PL intensity acquired in the $(730 \pm 8)$ nm spectral window from regions of samples #1 and #2 implanted at fluences of $1\times10^{16}$ cm$^{-2}$. The experimental data are plotted together with the relevant fitting curves with Eq. 2 by adopting $c = ½$. The different decay trends are clearly distinguishable, with $k$ values corresponding to $(0.65 \pm 0.02)$ and $(1.80 \pm 0.03)$ for samples #1 and #2, respectively. The corresponding quantum efficiency values are 58% and 27%, respectively. For sake of comparison, an ideal curve corresponding to no luminescence quenching (i.e. $k = 0$, $\eta = 100\%$) is also reported with the same background signal (blue line). The quality of the data fitting is very satisfactory, with reduced $\chi^2$ values of 1.3 and 1.5, respectively, thus confirming the validity of the previously described model. A fit with Eq. 2 using the dipolequadrupole coupling (i.e. $c = ⅜$) was less satisfactory for both NV$^0$ and NV$^-$ emissions, as reported in previous works [47]. This result, combined with the fact that both NV centers and single substitutional nitrogen have electric dipole, supports the hypothesis of dipole-dipole coupling.

With regards to the data analysis procedure, it is worth remarking that in all fitting procedures the uncertainties on the photon counts have been estimated according to Poissonian statistics, and that the temporal response function of the instrument has been suitably taken into account. As for the radiative lifetime constants, we assumed that in our optical excitation conditions (laser pulse wavelength, intensity and duration) an efficient initialization of the NV$^-$ centers in the $m_S = 0$ state [60] was obtained, despite the fact that optically-induced spin polarization in NV$^-$ centers is indeed limited to $\simeq 80\%$ [45, 61, 62]. This approximation was motivated by the fact that the resonant coupling strengths for the different types of transitions are unknown, as much as the exact degree of spin polarization of the defects in our system, and introducing all of the above-mentioned quantities as free parameters would have determined an excessive degree of mutual correlations in the fitting procedure. However, it was observed that the degree of uncertainty introduced by assuming a full $m_S = 0$ spin polarization of the NV$^-$ system is of the same order of magnitude (i.e. $\simeq 5 \div 10\%$) as the reported uncertainties on the resulting $k$ values. Therefore, in all fitting procedures a $(12 \pm 1)$ ns lifetime corresponding to the $m_S = 0$ spin state was set for the NV$^-$ decays [46]. The radiative lifetime of NV$^0$ centers was preliminarily estimated as $(17 \pm 1)$ ns in an optical grade CVD diamond from Element Six that was implanted with 10 MeV N ions at a fluence of $2 \times 10^{13}$ cm$^{-2}$ and subsequently annealed with the same procedure described above. As reported in Fig. 7, the decay was purely exponential within experimental uncertainty, therefore we conclude that in the above-mentioned sample the luminescent centers could be considered (within the sensitivity limit of our technique) as free from quenching effects. The obtained value represents one of the few estimations of radiative lifetime in $NV^0$ centers available in literature, in satisfactory agreement with previous results [47].

*4a. Frontal implantations*

As mentioned above, for samples #1 and #2 optical probing was performed in the same geometry as the ion implantations, i.e. perpendicularly to the main surface of the sample. This approach resulted in an optical measure arising from the integration along the implantation depth over a thickness corresponding to the probing depth of the technique (~10 μm). The implantation of areas at different fluences in the two samples allowed a systematic study of the variation of the NV lifetimes as a function of induced damage densities, which were parameterized in a simple linear approximation (i.e. by ignoring non-linear damage effects such as defect-defect interactions and self-annealing processes) by multiplying the vacancy linear density profile resulting from SRIM simulations (see Fig. 3) by the implantation fluence, thus obtaining a "parametric" vacancy density. Moreover, it is worth stressing that the reported vacancy density estimations are relevant to the samples prior to thermal annealing, therefore for this additional reason they cannot be accounted as a realistic estimation of the physical vacancy density in the samples under analysis. For each implantation fluence, a mean vacancy density was estimated by averaging the vacancy density profile across the probing depth of the technique.

Figs. 8a and 8b show the evolution of *k*, as evaluated from the fitting of relevant decay chronograms (see Fig. 6b), for the NV$^0$ and NV$^-$ emissions (respectively) from sample #1 as a function of 2 MeV H$^+$ implantation fluence. The above-mentioned mean vacancy density is also reported in the upper horizontal axis, while corresponding quantum efficiency values (as evaluated from the value of *k* with Eq. 3) are reported on the right-hand vertical axis in a nonlinear scale. In both cases, the variation of the *k* parameter follows a non-monotonic trend as a function of damage density.

As for the NV$^0$ emission data shown in Fig. 8a, the reported "*k* vs fluence" trend can be interpreted as follows. The initial decrease in the *k* value at low fluences is due to the fact that a significant fraction of the induced vacancies is recombining with native nitrogen atoms, thus increasing the population of NV centers in the sample. Therefore, it is reasonable to assume that, at increasing fluences, the NV$^0$ centers in the sample experience an atomic environment which is richer of NV centers and less rich of nitrogen atoms. While from previous reports the NV$^0$ centers have proved to be resonantly coupled with nitrogen atoms [47], the data seem to indicate that the resonant coupling from a NV$^0$ to a nearby quenching NV$^-$ center is not be very effective. Thus, at least in a small fluence regime, it is reasonable to assume that in lownitrogen samples the NV$^0$ centers experience a smaller quenching with increasing fluences. At higher fluences, we can assume that the formation process of NV centers starts saturating due to the limiting nitrogen concentration: therefore, additional ion-induced defects (vacancy-interstitial pairs and more complex defects) increasingly contribute to the quenching the NV$^0$ emission, thus giving direct evidence that the dipole-dipole energy transfer process between NV$^0$ centers and ion-induced defects is indeed effective.

The NV⁻ emission dataset shown in Fig. 8b is not particularly extensive because only at low fluences it was possible to reliably detect the NV⁻ emission intensity, while at higher fluences the phononic tail of the NV⁰ emission tends to overlap with the NV⁻ emission in the spectral detection range (see Fig. 4a). In this rather restricted fluence range, no significant variation of $k$ is measured. This can be explained as due to a weaker coupling of NV⁻ centers to substitutional nitrogen atoms with respect to NV⁰ centers (see the typical $k$ values in Figs. 8a and 8b, for comparison), as well as to weaker coupling with ion-induced structural defects (at least in the restricted fluences range reported here).

Similarly to what reported in Fig. 8, Fig. 9 shows the evolution of $k$ for the NV⁻ emission from sample #2 as a function of 2 MeV H⁺ implantation fluence. Also in this case, mean vacancy density and quantum efficiency values are also reported on additional axes. It is worth stressing that, given the higher nitrogen concentration, in sample #2 no reliable signal could be acquired from the NV⁰ emission, and therefore (differently to what observed in sample #1) the NV⁻ emission could be reliably detected at higher fluences. From Fig. 9 it is clear how the $k$ value is basically unaffected from implantation conditions, as long as the fluence values are below ~$1\times10^{16}$ cm⁻² (i.e. average vacancy densities of ~60 ppm); this is compatible with the lowfluence regime observed for NV⁻ emission also in sample #1 (see Fig. 8b). When damage levels exceed the above-mentioned threshold, $k$ starts increasing with a logarithmic dependence from the fluence. The same type of variation of $k$ is also observed as a function of average damage density, according to the above-mentioned linear approximation. By fitting both the semilogarithmic increase and the plateau at low damage densities due to the background nitrogen concentration, the following relation is obtained:

$$k(v) = \begin{cases} (1.803 \pm 0.012) \, for \, v < 60 \, ppm \\ (0.411 \pm 0.012) + (0.34 \pm 0.06) \cdot \ln(v[ppm]) \, for \, v > 60 \, ppm \end{cases} \quad (4)$$

The observed trends are interpreted as follows: at low fluence values the dominating factor in determining a variation in the NV⁻ centers lifetime is represented by the concentration of native substitutional nitrogen, while at increasing implantation fluences the ion-induced damage is the key factor in decreasing the centers lifetime. The significant correlation between estimated vacancy densities and $k$ seem to indicate that indeed, at first approximation, isolate vacancyinterstitial defects are the main factor determining NV⁻ centers quenching, although the contribution of more complex defects cannot be ruled out.

*4b. Lateral implantations*

As mentioned above, for samples #3 optical probing was performed in an orthogonal geometry with respect to the ion implantations, i.e. parallel direction to the main implanted surface of the sample (see Fig. 2). This approach resulted in an optical measurement which was directly following the implantation depth, while integrating the signal along a constant vacancy density.

Fig. 10 shows a typical TCSPC map acquired in the $(730 \pm 8)$ nm spectral window (corresponding to the phonon sideband of the NV⁻ emission) from a region of sample #3 implanted at a fluence of $1 \times 10^{16}$ cm$^{-2}$. As reported for sample #2, due to the high nitrogen concentration no significant NV⁰ emission could be detected in sample #3. The position of the sample surface is highlighted in the figure, and the depth direction is towards the right side of the map. The micrograph is superimposed with the semi-logarithmic vacancy density profile reported in Fig. 3 for qualitative comparison of SRIM simulations and experimental data. As expected, Fig. 10 clearly shows how the photoluminescence yield increases at increasing depths in the sample bulk, up to the maximum intensity corresponding to the ions end-of-range depth (~25 µm). This is clearly due to the fact that higher vacancy densities in the damage profile correspond to higher NV centers concentrations, and therefore to stronger PL intensities.

As for frontal implantations, lifetime chronograms have been extracted from the lateral map on a pixel-by-pixel basis and the resulting fitting $k$ parameters have been averaged across constant depths in the sample (i.e. along vertical lines in the map reported in Fig. 10). The resulting depth profiles of the $k$ values are reported in Fig. 11a. As for Figs. 8 and 9, also in this plot the corresponding values of the quantum efficiency (as evaluated with Eq. 3) are reported on the right vertical axis in a nonlinear scale.

Apart from the direct measurement in lateral geometry, the depth profile of the $k$ value can also be derived from the vacancy density profile obtained from SRIM simulations by applying a variation of Eq. 4 in which the plateau value of $k$ for low damage densities is lower due to the slightly lower nitrogen concentration in this sample as compared to sample #2 (i.e. 190 ppm instead of 200 ppm). Such plateau value ($k = 1.6$) was derived for sample #3 from TCSPC measurements in frontal geometry on areas irradiated at the lowest fluences (see the following section for further details). As shown in Fig. 11a, the directly-measured $k$ depth profiles are in good agreement with the above-mentioned numerical predictions (with the possible exception of the data at the very end of range of implanted ions where the agreement is less satisfactory), thus confirming that the linear approximation in the estimation of vacancy density and the relation reported in Eq. 4 provide an adequate description of the PL quenching process, at least in the damage density ranges reported here.

The measured $k$ values can be correlated with the SRIM-derived estimations of the average vacancy density established at different depths in the implanted samples. The resulting plot is reported semi-logarithmic scale in Fig. 11b: the observed quasi-linear dependence is in good agreement with what reported in Fig. 9 for frontal implantations. Here, the mismatch between experimental and SRIM data at the ions end of range is reflected in the deviation from linearity at high damage densities.

*4c. Low-fluence implantations - Effect of nitrogen concentration*

The results of frontal TCSPC measurements of the NV$^-$ emission from the regions implanted at the lowest fluences (i.e. $F < 3 \times 10^{15}$ cm$^{-2}$) in all three samples were mutually compared, with the scope of elucidating the effect of native nitrogen on non-radiative coupling in a regime in which the effect of ion-induced damage is negligible (see Figs. 8 and 9). A similar study was not performed for the NV$^0$ emission since a significant native NV$^0$ emission was only measurable in sample #1. Fig. 12 shows the variation of the $k$ parameter derived from the fitting of the chronograms acquired in the $(730 \pm 8)$ nm spectral range as a function of the substitutional nitrogen concentration estimated with the FTIR measurements (see Section 2a). The corresponding quantum efficiency, as derived from $k$ through Eq. 3, is also reported on the right vertical axis. The data shown in Fig. 12 are compatible within the reported uncertainties with a linear correlation between $k$ and $[N_S]$. The resulting fitting curve for $k$ vs $[N_S]$ is:

$$k = (2 \pm 1) \times 10^{-1} + (8 \pm 1) \times 10^{-3} \cdot [N_S] \qquad (5)$$

It is worth noting that in the result above a significant contribution to the uncertainty of the obtained parameters is propagated from the uncertainty on the value of the radiative lifetime. Having said that, it can be observed that, since the intercept value in the linear regression is affected by a significant uncertainty, it is not possible to rule out that a non-zero $k$ value is observed in an ideal sample with no substitutional nitrogen, and therefore that other kinds of native defects affect the NV$^-$ lifetime. It is worth mentioning that for $[N_S] \cong 100$ ppm (a value which is typical for type Ib samples produced with the HPHT technique) the linear fit yields a $k$ value corresponding to a quantum efficiency of $\eta \cong 50\%$ through the application of Eq. 3. Undoubtedly, this kind of analysis would benefit from the characterization of a larger set of samples characterized by a broader spectrum of $[N_S]$, but this goes beyond the scope of the present work.

## 5. Conclusions

A time-correlated single photon counting (TCSPC) microscopy technique was successfully employed to characterize the lifetime of NV$^-$ and NV$^0$ centers in type Ib HPHT single-crystal samples with different concentrations of native substitutional nitrogen, after systematic 2 MeV H$^+$ ion implantation in different geometries and at increasing fluences. The obtained results can be summarized as follows:

- while a radiative lifetime for NV$^-$ centers of (12 ± 1) ns was taken from literature [46], the radiative lifetime for NV$^0$ centers was directly measured in an opticalgrade Nimplanted sample, yielding a value of (17 ± 1) ns, in satisfactory agreement with literature [47];

- the dipole-dipole resonant energy transfer was successfully applied to model the measured decay times of NV$^0$ and NV$^-$ centers in different experimental conditions, allowing the determination of empirical parameter *k* accounting for the strength of the non-radiative coupling between the PL centers under investigation and native/induced defects in the samples;

- in frontally-implanted samples, the variation of *k* parameter was reported as a function of ion-induced defect density for both NV$^0$ and NV$^-$ emissions: while a non-monotonic variation was observed for NV$^0$ emission, a logarithmic increase of *k* vs damage density was found for NV$^-$ emission for damage levels exceeding a critical threshold determined by the background nitrogen concentration; the data relevant to NV$^-$ emission were quantitatively analyzed;

- in laterally-implanted samples, it was possible to map the evolution of the PL lifetime across the sample thickness, thus obtaining a direct evidence of the effects of the strongly non-uniform damage profile; the experimental data exhibited a good agreement with numerical prediction based on SRIM simulations combined with the results of the quantitative analysis carried out for frontally implanted samples;

- in regions implanted at the lowest fluences, a linear dependence was identified between native nitrogen substitutional concentration independently measured by FTIR spectroscopy and the *k* value; this allowed the attribution of the quenching of NV$^-$ emission primarily to nitrogen, although other kind of native defects cannot be ruled out in principle.

Finally, it is worth noting that the evaluation of the empirical parameter *k* allows the estimation of the mean quantum efficiency of the luminescence centers, which is a physical property of extreme relevance for the current studies on the employment of diamond color centers for photonic applications, particularly with regards to those fabrication/functionalization processes involving ion beam irradiation.

**Acknowledgments**

We thank J. Meijer for the 10 MeV N implantations performed at the "heavy ion leg" beam of the 4 MV Tandem accelerator of the RUBION laboratories (Ruhr Universität Bochum). This work is supported by the following projects and grants, which are gratefully acknowledged: FIRB "Future in Research 2010" project (CUP code: D11J11000450001) funded by the Italian Ministry for Teaching, University and Research (MIUR); "Dia.Fab." experiment at the INFN Legnaro National Laboratories; one-month DAAD 2011 grant "Ion implantation in diamond for applications in photonics" funded by the German Academic Exchange Service; University of Torino-Compagnia di San Paolo projects "OLIPATEN12, 2012 - Call 1" and "ORTO11RRT5, 2011 - Linea 1A".

**Figures**

Fig. 1

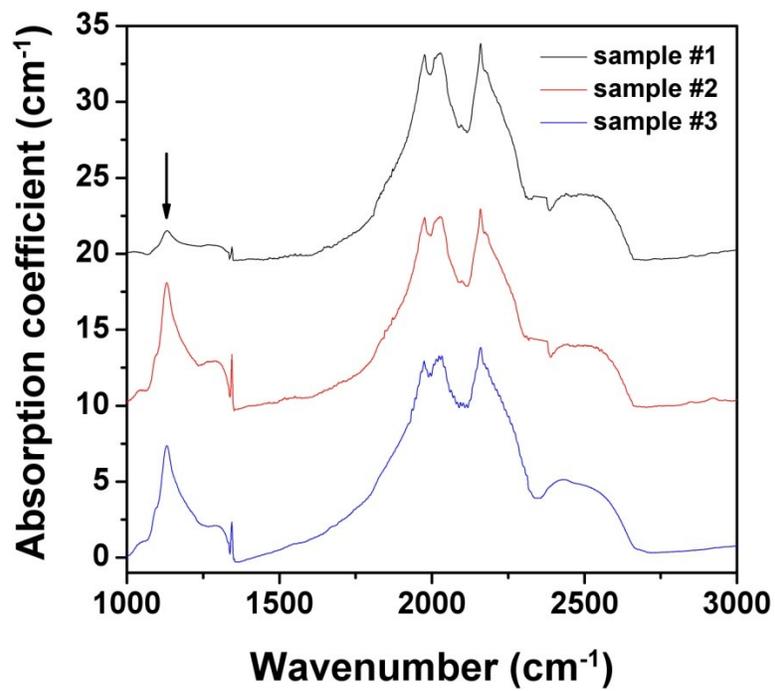

Fig. 2

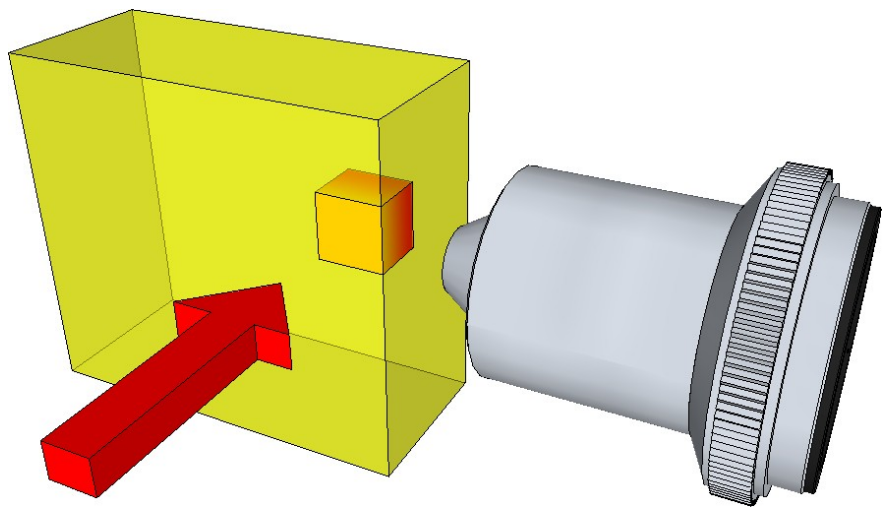

Fig. 3

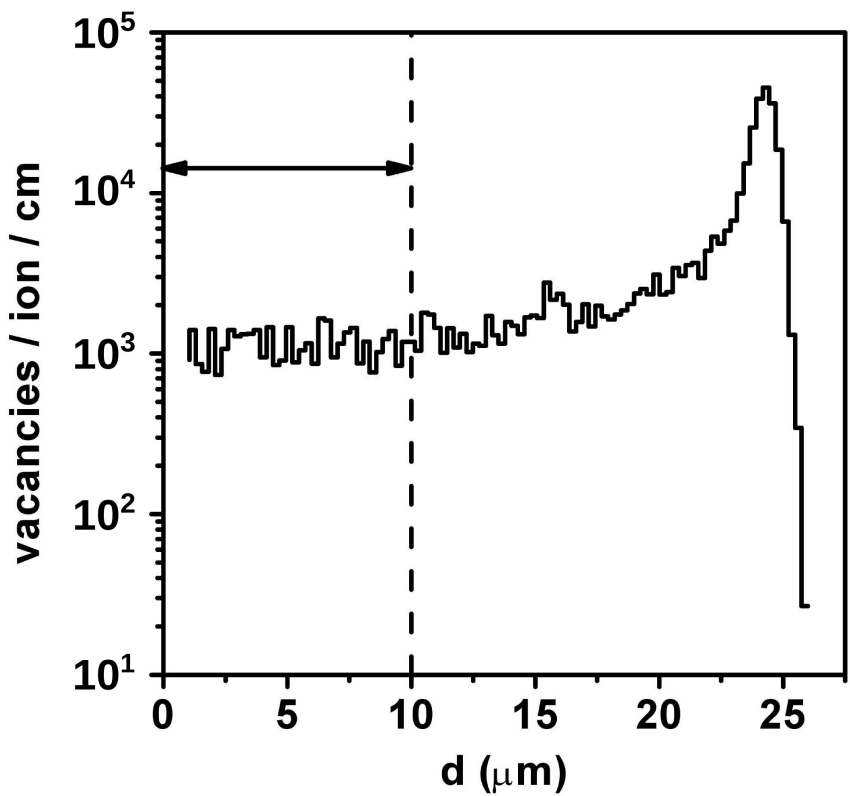

Fig. 4

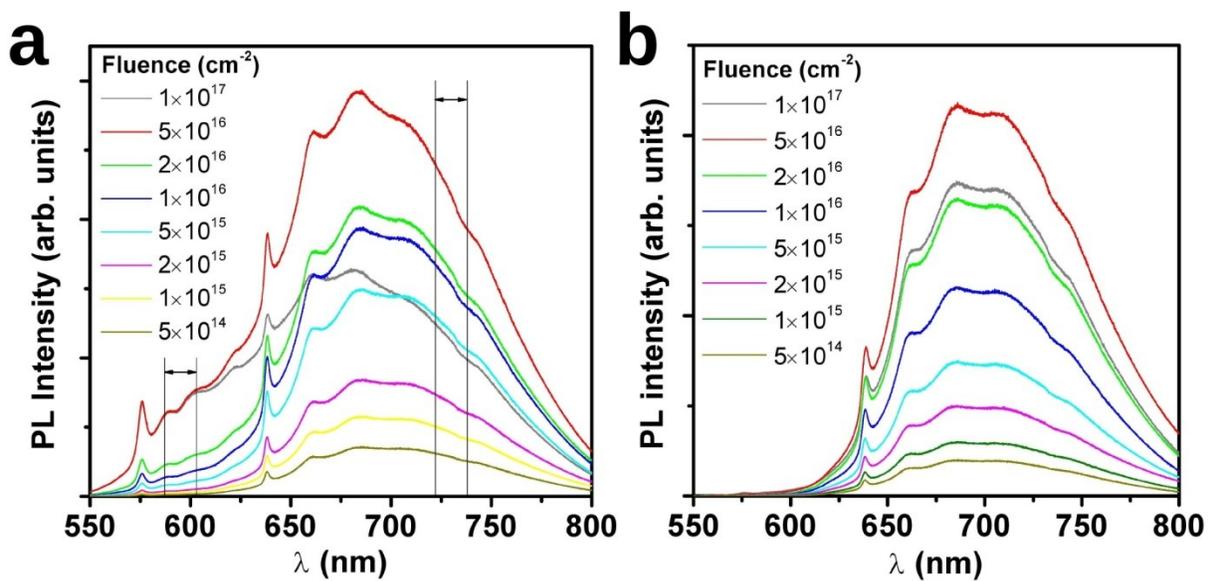

Fig. 5

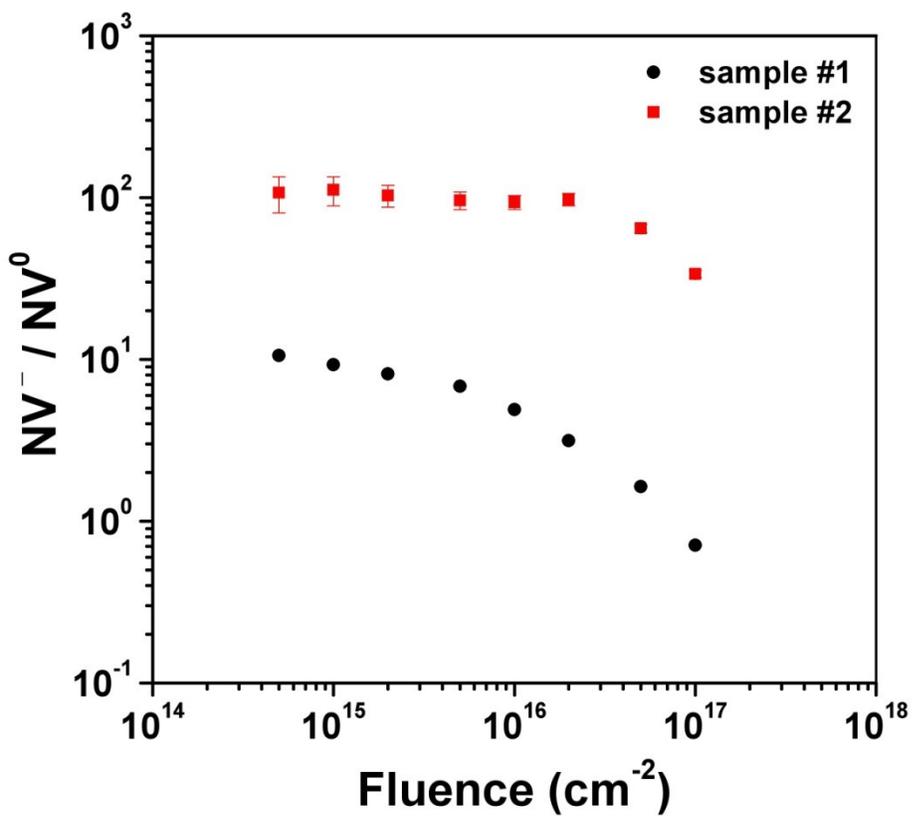

Fig. 6

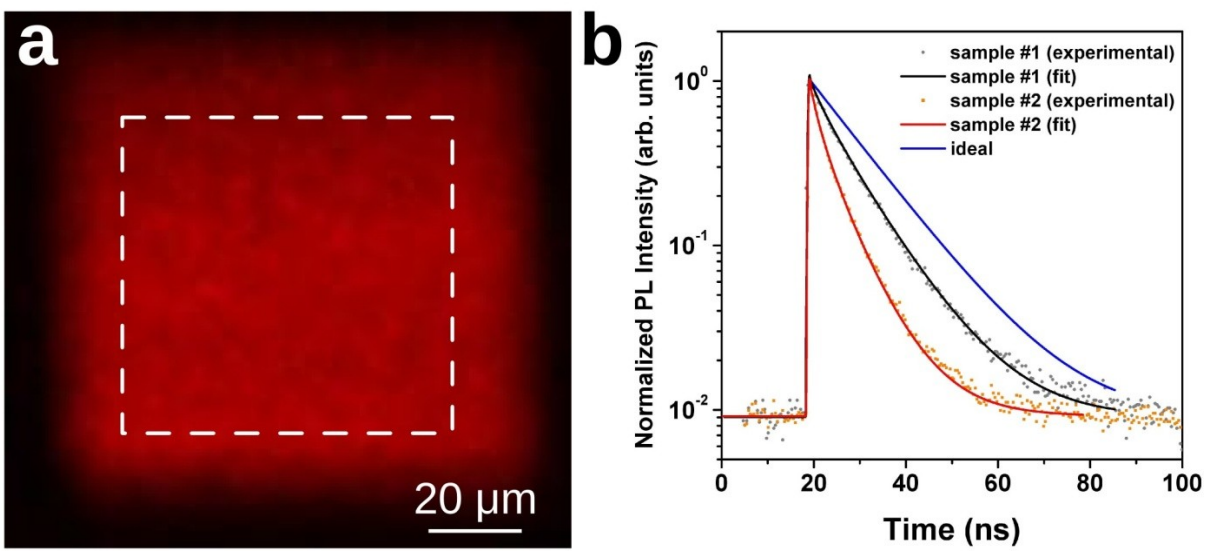

Fig. 7

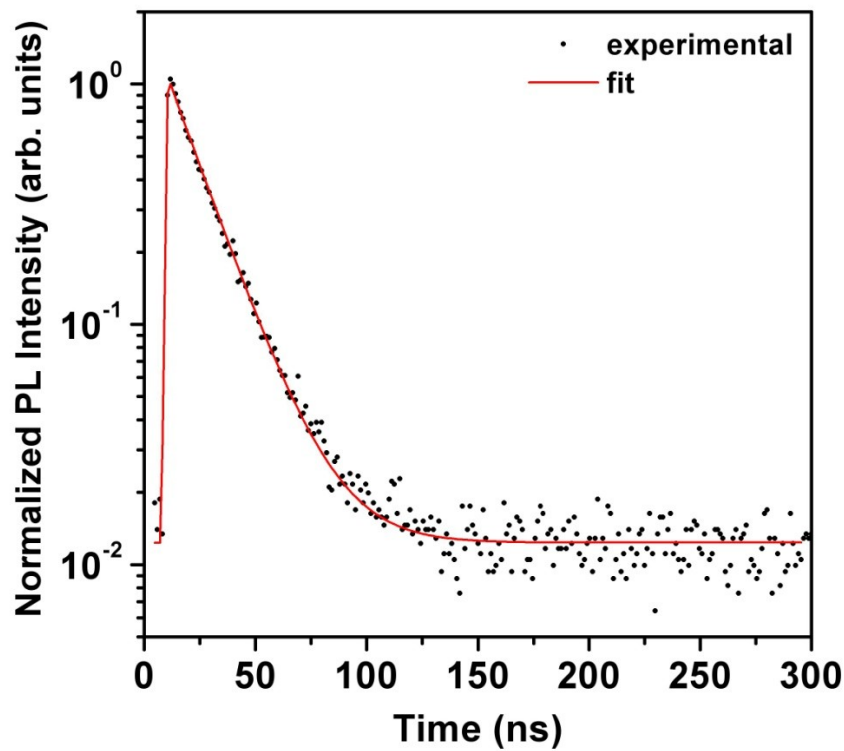

Fig. 8

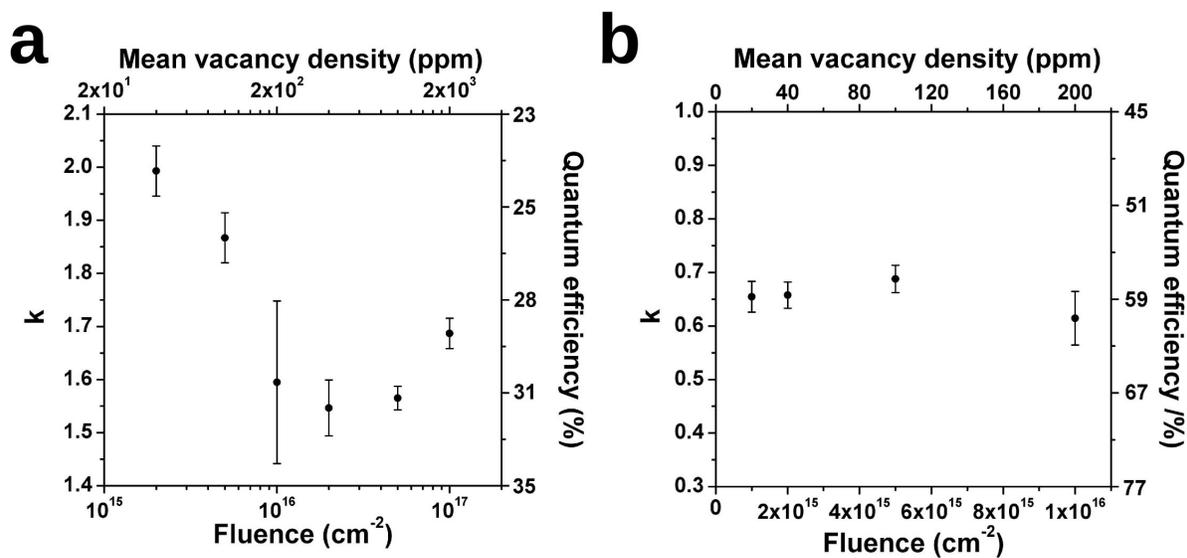

Fig. 9

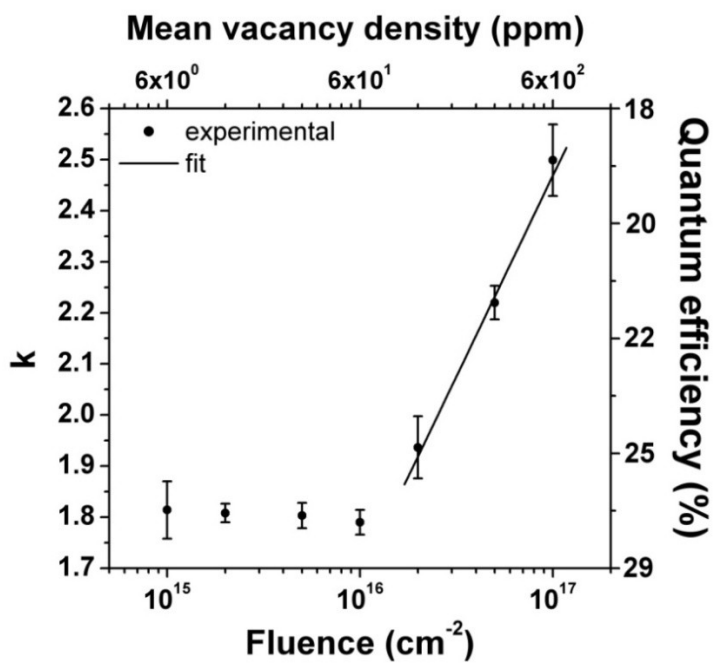

Fig. 10

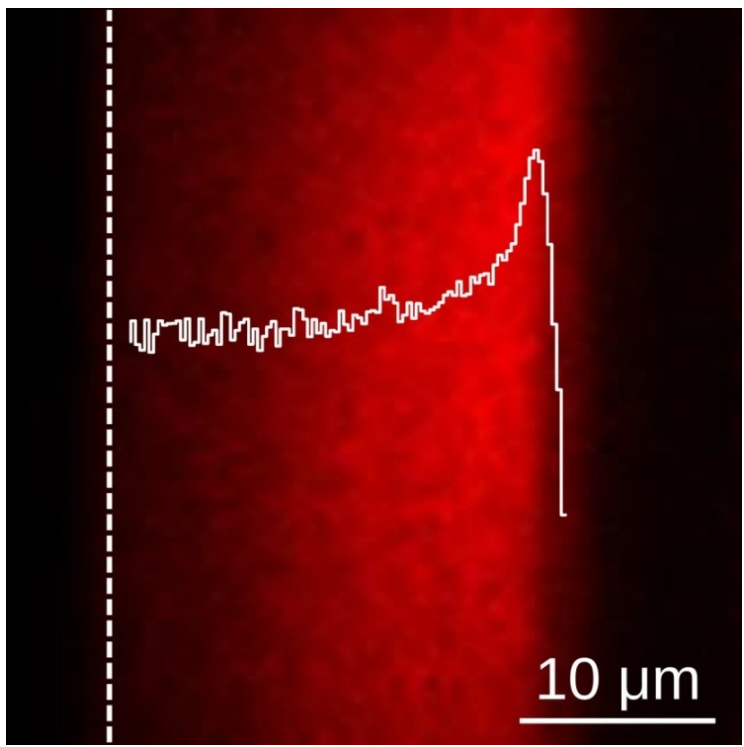

Fig. 11

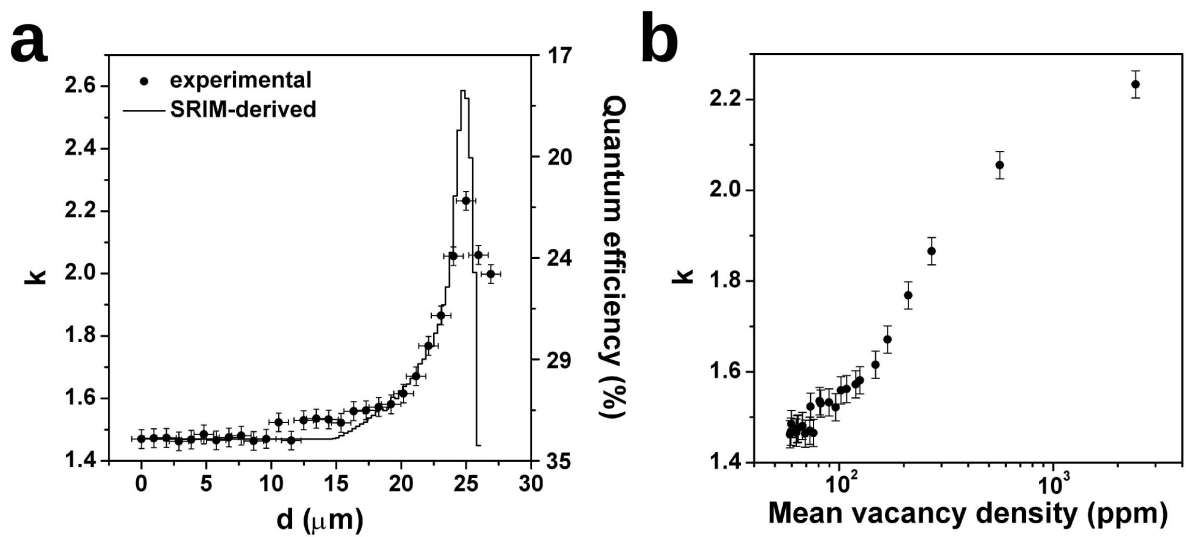

Fig. 12

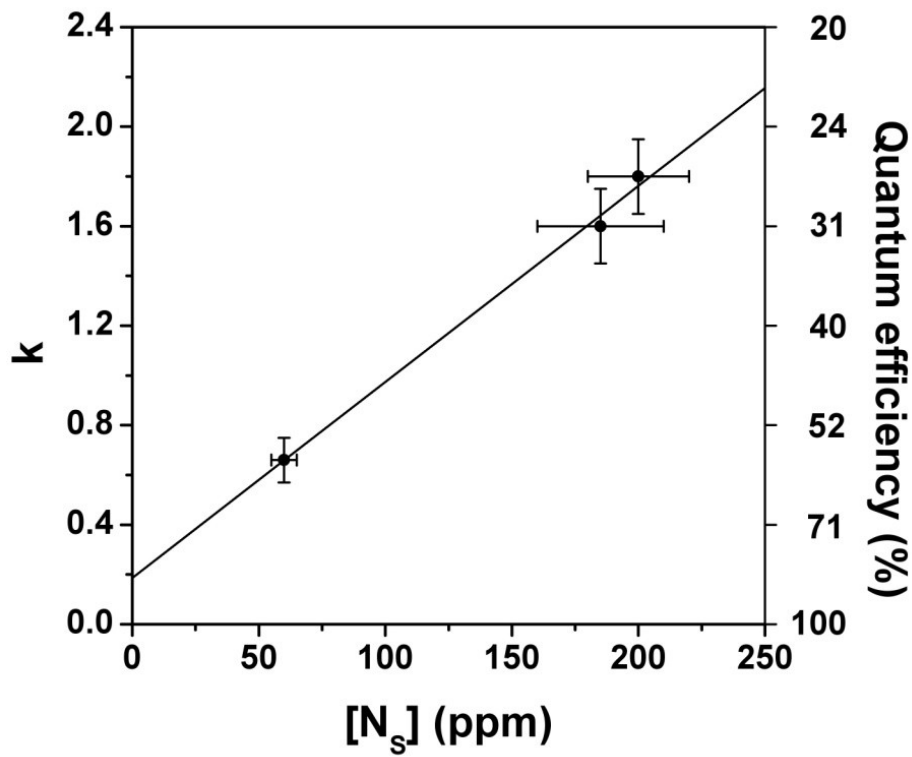

**Figure captions**

Fig. 1 (color online): Micro-FTIR spectra from samples #1 (black line), #2 (red line) and #3 (blue line) before ion implantation. The spectra from samples #1 and #2 are displaced along the vertical axis by 10 cm$^{-1}$ and 20 cm$^{-1}$ for sake of clarity. The spectrum from sample #3 is correctly referred to the vertical axis in terms of absolute values. The black arrow indicates the 1130 cm$^{-1}$ feature associated to the absorption from single substitutional nitrogen [51].

Fig. 2 (color online): schematic representation of the ion implantation performed on sample #3 with the purpose of allowing cross-sectional optical characterization in a lateral geometry. The yellow parallelogram represents the diamond sample, with one of its two main faces frontally exposed to ion implantation; the red arrow indicates the direction of the incoming ion beam which is irradiating a square area (highlighted in orange within the sample) across the sample edge. The optical objective on the right side represents the optical setup which is scanning the sample orthogonally with respect to the irradiation direction. The drawing is not to scale.

Fig. 3: depth profile of the damage linear density, as evaluated from SRIM2008.04 Monte Carlo code [52]. The damage density is parameterized in number of vacancies per incoming ion and unit length in the depth direction. The vertical dashed line highlights the probing depth of the TCSPC microscopy technique.

Fig. 4 (color online): Room-temperature PL spectra from samples #1 (a) and #2 (b) after ion implantation and thermal annealing. Different spectra correspond to different implantation fluences, as indicated in the legends. Both NV$^-$ (ZPL: $\lambda = 638$ nm) and NV$^0$ (ZPL: $\lambda = 575$ nm) emissions are visible, while no GR1 (ZPLs: $\lambda = 740.9$ nm, 744.4 nm) emission is observed. The $(730 \pm 8)$ nm and $(595 \pm 8)$ nm spectral intervals adopted in TCSPC measurements are highlighted in (a).

Fig. 5 (color online): evolution of the (NV$^-$/NV$^0$) ZPL emission ratio as a function of implantation fluence, for samples #1 (black circles) and #2 (red squares). In the data relevant to sample #1 the uncertainty bars are smaller than the symbol size and have therefore not been reported.

Fig. 6 (color online): (a) Typical TCSPC map acquired in the $(730 \pm 8)$ nm spectral window (corresponding to the phonon sideband of the NV$^-$ emission) from a 100×100 μm$^2$ region of sample #2 implanted with 2 MeV H$^+$ ions at a fluence of $1 \times 10^{16}$ cm$^{-2}$; the region from which chronograms are extracted and analyzed is highlighted by the white dashed square. (b) Typical chronograms of normalized PL intensity acquired in the $(730 \pm 8)$ nm spectral window from a region of sample #1 implanted at a fluence of $1 \times 10^{16}$ cm$^{-2}$ (gray circles: experimental data, black line: fit), sample #2 implanted at a fluence of $1 \times 10^{16}$ cm$^{-2}$ (orange squares: experimental data, red line: fit); an ideal curve corresponding to no luminescence quenching (i.e. $k = 0$) is also reported for comparison (blue line).

Fig. 7 (color online): Chronogram of normalized $NV^0$ PL intensity acquired in the (595 ± 8) nm spectral window from an optical grade CVD diamond sample after 10 MeV N implantation at $2\times10^{13}$ cm$^{-2}$ fluence and subsequent 800 °C annealing. Experimental data (black circles) are reported with an exponential fitting curve (red line) from which a lifetime of (17 ± 1) ns is resulting.

Fig. 8: evolution of $k$ for the $NV^0$ (a) and $NV^-$ (b) emissions from sample #1 as a function of 2 MeV $H^+$ implantation fluence. The two graphs are reported with the same extension (i.e. $\Delta k = 0.7$) in the left vertical axis, for comparison. The mean vacancy density across the technique probing depth is reported in the upper horizontal axis, while corresponding quantum efficiency values (as derived from Eq. 3) are reported on the right-hand vertical axis in a non-linear scale.

Fig. 9: evolution of $k$ for the $NV^-$ emission from sample #2 as a function of 2 MeV $H^+$ implantation fluence. The mean vacancy density across the technique probing depth is reported in the upper horizontal axis, while corresponding quantum efficiency values (as derived from Eq. 3) are reported on the right-hand vertical axis in a non-linear scale. Experimental data (dots) are reported together with linear fit of the last 4 points (line).

Fig. 10 (color online): typical TCSPC map acquired in lateral geometry (see Fig. 2) within the (730 ± 8) nm spectral window (corresponding to the phonon sideband of the $NV^-$ emission) from sample #3. The position of the sample surface is highlighted by the vertical dashed line, and increasing depths in the sample bulk span in the horizontal direction towards the right side. The continuous line profile highlights the linear vacancy density profile in semilogarithmic scale reported in Fig. 3 for comparison of SRIM simulations and experimental data.

Fig. 11: $k$ values from a region which was laterally implanted at a fluence of $1 \times 10^{16}$ cm$^{-2}$ in sample #3. a) $k$ as a function of depth from the sample surface: (dots) experimental data resulting from the chronograms encoded in the TCSPC map reported in Fig. 10 after averaging at constant depths within the sample (i.e. along the vertical axis in Fig. 10); (line) numerical prediction based on SRIM output and Eq. 4; quantum efficiency values (as derived from Eq. 3) are displayed on the right vertical axis in a non-linear scale. b) $k$ as a function of the mean vacancy density at different depths as evaluated with SRIM simulations.

Fig. 12: variation of the $k$ parameter estimated from the fitting of the chronograms acquired in the $(730 \pm 8)$ nm spectral range (corresponding to the phonon sideband of the NV$^-$ emission) from samples #1, #2 and #3 prior to ion implantation, as a function of substitutional nitrogen concentration estimated with FTIR measurements. On the right vertical axis the corresponding quantum efficiency values (see Eq. 3) are reported. Experimental data with uncertainties (dots) are reported with the linear fitting curve (line).